\documentclass[12pt]{article}
\usepackage{amsmath,amsfonts,xypic, epsfig}
\usepackage{amssymb}
\usepackage{amscd}
\usepackage[all]{xy}
\advance \textheight by \topskip
\makeatletter
\@addtoreset{equation}{section}
 \def\theequation{\thesection.\arabic{equation}}
\makeatother
\def\NN{{\mathbb N}}

\def\NN{{\mathbb N}}

\newcommand{\beqa}{\begin{eqnarray}}
\newcommand{\eeqa}{\end{eqnarray}}
\newcommand{\noi}{\noindent}

\def\>{\rangle}
\def\<{\langle}

\begin{document}

\title{
%\begin{flushright}
%{\small USACH-FM-01-02}\\[-0.4cm]
%{\small PM-01-07}\\[1cm]
%\end{flushright}
{\bf Commutative limit of a renormalizable noncommutative  model
}  
\author{ 
{\sf   
Jacques Magnen${}^{a}$\footnote{jacques.magnen@cpht.polytechnique.fr}, Vincent Rivasseau${}^{b}$\footnote{vincent.rivasseau@th.u-psud.fr} and Adrian Tanasa${}^{a,c}$\footnote{adrian.tanasa@ens-lyon.org}
}
\\
{\small ${}^{a}${\it Centre de Physique Th\'eorique, CNRS UMR 7644,}} \\
{\small {\it Ecole Polytechnique, 91128 Palaiseau, France}}  \\ 
{\small ${}^{b}${\it Laboratoire de Physique Th\'eorique, CNRS UMR 8627,}} \\
{\small {\it b\^at. 210, 
Universit\'e Paris XI, 91405 Orsay Cedex, France}}  \\ 
{\small ${}^{c}${\it Institutul de Fizica si Inginerie Nucleara Horia Hulubei,}} \\
{\small {\it P. O. Box MG-6, 077125 Bucuresti-Magurele, Romania}}  \\ 
}}
%\date{\today}
\maketitle
\vskip-1.5cm

\vspace{2truecm}

\begin{abstract}
\noindent
Renormalizable $\phi^{\star 4}_4$ models on Moyal space have been 
obtained by modifying the commutative propagator. But these models have a divergent ``naive'' commutative limit.
We explain here how to obtain a coherent such commutative limit for a recently proposed  
translation-invariant model. The mechanism relies on the analysis of the uv/ir mixing in Feynman graphs at any order in perturbation theory.
\end{abstract}


\newpage

\section{Introduction}
%; commutative limit of renormalizable non-commutative models}
\renewcommand{\theequation}{\thesection.\arabic{equation}}    
\setcounter{equation}{0}

Noncommutative field theory \cite{snyder} is a subject which lies at the intersection
of many different perspectives. It is a natural generalization of Alain Connes 
noncommutative geometry program \cite{connes},
it is  an effective regime of string theory \cite{string1,string2}, and of loop quantum gravity \cite{Freidel:2005me}.
It has also the potential to throw light on difficult non perturbative
physical problems (quantum Hall effect \cite{hall1, hall2, hall3}, quark confinement...). 
For a recent review, see \cite{QS}.

The issue of renormalization of noncommutative (Euclidean) 
field theory on Moyal space however was made difficult
by the discovery of ultraviolet/infrared mixing \cite{melange}. 
In the simple case of the $\phi^{\star 4}_4$ model a first solution was 
obtained by H. Grosse and R. Wulkenhaar, who modified the ordinary propagator, adding
a harmonic potential term \cite{GW}. The resulting theory 
has a new symmetry called Langmann-Szabo or LS duality \cite{LaSz}. 
This result has sparked a flurry of activity. The corresponding "GW" model (and other related models) have been shown
fully renormalizable by all the main methods; furthermore, different field theoretical properties have been exbited \cite{RVW,GMRV,param1,param2,mellin,dimreg,hopf,goldstone}. Even more important, 
it is the first example of a non supersymmetric field
theory in four dimensions which has a non trivial ultraviolet fixed point \cite{beta1,beta23,beta}.
It is currently under construction in a non perturbative sense \cite{Riv2, MR}, something which has never been fully done for any four dimensional commutative theory. Noncommutative gauge theories 
with some version of LS symmetry are actively searched for \cite{gauge1,gauge2,gauge3}.
However in spite of this great conceptual and mathematical interest 
the GW model has for the moment no direct  phenomenological applications. 
It breaks translation invariance and it seems very difficult to connect it 
continuously to ordinary field theory at lower energy.
In technical terms the $\theta \to 0$ limit of the GW model seems too singular.

These drawbacks motivated the  
introduction in  \cite{noi}  of another $\phi^{\star 4} _4$ model which is translation invariant and also renormalizable to all orders in perturbation theory. It is not based on the LS symmetry, hence it 
may not be fully constructible in a non perturbative sense. However it may be easier to 
connect to ordinary high energy physics. To support this idea comes the recent proof \cite{beta-GMRT} that the $\beta$ function of this model is just a rational multiple of the commutative model $\beta$ function. Let us emphasize here that this result, in spite of the title of \cite{beta-GMRT}, was proven there at {\it any order in perturbation theory}.

Let us further state here that the parametric representation for this model was obtained in \cite{param-GMRT}. Also some one-loop and higher order Feynman amplitudes were explicitly computed in \cite{ei}. Furthermore, the static potential associated to this noncommutative model was calculated in \cite{static}. 
Moreover, the idea of the translation-invariant scalar model proposed in \cite{noi} was also extended to the level of noncommutative gauge fields \cite{gauge-GMRT}. For a review of this developments, the interested reader may report himself to \cite{rev-io}.

In this paper we study how in this model  the $1/p^2$ counterterms created by the uv/ir mixing graphs
can morph  into the ordinary mass and wave function counterterms that  these graphs 
generate in the ordinary commutative $\phi^4_4$  as the $\theta$ 
parameter is turned off. In this way we show how a renormalizable
noncommutative model can become an effective commutative model. This is
a small step in the direction of smoothing the road between commutative and noncommutative
field theory. 
Note that such a mechanism could also give insights on the noncommutative limit of the gauge models \cite{gauge-GMRT}.
Commutative field theory certainly works well at least up to the LHC energies,
but a noncommutative field theory regime might be relevant
somewhere between the LHC scale and the Planck scale where gravity 
has to be quantized.

\section{Scalar field theory on the Moyal space}
\renewcommand{\theequation}{\thesection.\arabic{equation}}   
\setcounter{equation}{0}

In this section we recall the definition of the noncommutative Moyal space on which we consider scalar quantum field theory. Furthermore, some general considerations which respect to the associated Feynman graphs are given.

\subsection{ The ``naive''  $\phi^{\star 4}$ model}

This model is obtained by replacing in the ordinary $\phi^4$ action  the pointlike product  
 by the Moyal-Weyl $\star$-product
\beqa
\label{act-normala}
S[\phi]=\int d^4 x \Big(\frac 12 \partial_\mu \phi \star \partial_\mu \phi +\frac
12 m^2 \phi\star \phi  + \frac{\lambda}{4!} \phi \star \phi \star \phi \star \phi\Big)(x),
\eeqa
with Euclidean metric. The commutator of two coordinates is
\beqa
[x_\mu, x_\nu]_\star=i \Theta_{\mu \nu},
\eeqa
where
\beqa
\label{theta}
\Theta=
\begin{pmatrix}
   0 &\theta & 0 & 0\\   
   -\theta & 0  & 0 & 0\\
   0&0 & 0 & \theta\\  
   0& 0 & -\theta & 0  
  \end{pmatrix},
\eeqa
and $\theta$ is the noncommutativity parameter.
% with mass dimension -2. 
%The expression of this $\star$-product is
%\begin{align}
%(\phi_1\star\phi_2)(x)=\frac{1}{\pi^4\theta^4}\int d^4yd^4z\ \phi_1(x+y)\phi_2(x+z)e^{-2iy_\mu\Theta^{-1}_{\mu\nu}z_\nu}.
%\end{align}
In momentum space, the action \eqref{act-normala} becomes
\beqa
\label{act-normala-p}
S(\phi)=\frac{1}{(2\pi)^4}\int d^4 p \Big(\frac 12 p^2 \phi(-p)\phi(p) +\frac
12 m^2 \phi(-p)  \phi(p)  + \frac{\lambda}{4!}V_{\theta}(\phi,p)\Big),
\eeqa
where the interaction potential is:
\begin{align}
V_\theta(\phi,p)=\frac{1}{(2\pi)^8}\int d^4qd^4k\ \hat\phi(p)\hat\phi(q)\hat\phi(k)\hat\phi(-p-q-k)e^{-\frac i2(p\wedge q+p\wedge k+q\wedge k)}.
\end{align}
Note that we have used the notation:
$$p\wedge q=p_\mu\Theta^{\mu\nu}q_\nu.$$
%Notice that the propagator associated to \eqref{act-normala-p} is the same to the one in commutative theory. The only change is in the interaction part: $V_{int}\to V_{int}^\star$.

\subsection{Feynman graphs: planarity and non-planarity, rosettes}
\renewcommand{\theequation}{\thesection.\arabic{equation}}   
\setcounter{equation}{0}
\label{Feynman}

In this subsection we give some useful conventions and  definitions.
Consider a $\phi^{\star 4}$ connected graph with $n$ vertices, $L$ internal lines and $F$
faces. 
One has
\beqa
\label{genus}
2-2g  =n-L+F,
\eeqa
\noi
where $g\in\NN$ is the {\it genus} of the graph.
If $g=0$ the graph is {\it planar}, if $g>0$ it is {\it non-planar}. 
Furthermore, we call $B$ the number of faces broken by external lines, and we say
that a planar graph is {\it regular} if $B=1$. 

The $\phi^4$ graphs also obey the relation
\beqa
\label{alta}
L= 2n  -   N/2,
\eeqa
where $N$ is the number of external legs of the graph.

In \cite{filk}, several contractions or "Filk moves"  were  defined on a Feynman 
graph. The first Filk move, which is the one we use in the sequel,
consists in reducing a tree line by gluing up
together two vertices into a bigger one.

\begin{figure}
\centerline{\epsfig{figure=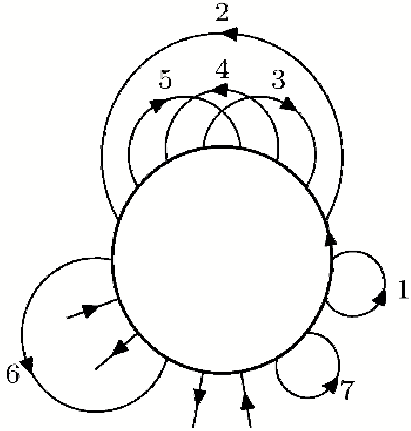,width=6cm} }
\caption{An example of a rosette}\label{roz}
\end{figure}

Repeating this operation for the $n-1$ lines of a spanning tree, one obtains a single final
vertex with all the loop lines hooked to it - a {\it rosette} (see Fig. \ref{roz}).

Note that the number of faces or the genus of the graph do not change 
under this operation. Furthermore, the external legs will break the same faces on the rosette. When one deals with a planar graph, there will be no crossing between the loop lines on the rosette. The example of Fig. \ref{roz} corresponds thus to a non-planar graph (one has crossings between the loop lines $3$, $4$ and $5$). 

In \cite{filk} the general oscillating factors appearing in the Feynman integrand as a function of the corresponding rosette were computed. We do not recall here these results, but will use
some particular cases when necessary.

\section{Uv/ir mixing as insight for the commutative limit} \label{2puncte}
\renewcommand{\theequation}{\thesection.\arabic{equation}}   
\setcounter{equation}{0}

The uv/ir mixing comes from the 2-point planar irregular graphs. We will first recall
the analysis of the ``non-planar'' tadpole and then extend it
to any 2-point planar irregular Feynman graph.

\subsection{The ``non-planar'' tadpole}

\begin{figure}[ht]
\centerline{\psfig{figure=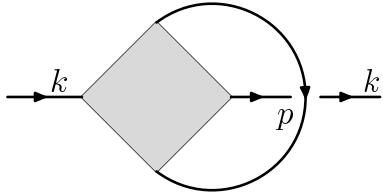,width=5cm}}
\caption{The ``non-planar'' tadpole graph.}
\label{fig:tadpole}
\end{figure}

The Feynman amplitude ${\cal A}_{G_0}$  of the ``non-planar'' tadpole $G_0$ of Fig. \ref{fig:tadpole} 
writes, up to some constant factor

\beqa
\label{ampli-tp}
{\cal A}_{G_{0}}(k) = \int d^4 p \, \frac{e^{i p\wedge k}}{p^2+m^2}.
\eeqa
The oscillating factor above is responsible for the convergence of the integral. One can thus interpret $\theta^{-\frac 12}$ as some kind of uv cut-off. Indeed, when  
%$\Theta^{\mu\nu}k_\nu= 0$, which corresponds to
%\beqa
$\theta=0$
% \mbox{ or } k=0,
%\eeqa
the integral is no longer convergent and \eqref{ampli-tp} will simply correspond to the divergent planar tadpole, leading to the usual {\it mass renormalization}. 
%Let us further detail this. 
This observation comes from the fact that the amplitude \eqref{ampli-tp} computes to
\beqa
\begin{cases}
				\infty & \mathrm{ if }~\theta=0,\\
				 \sqrt{\frac{m^2}{\theta^2 k^2}}K_1(\sqrt{m^2 \theta^2 k^2})\approx_{|p|\to 0} \frac{1}{\theta^2 k^2}& \mathrm{ if }~\theta \ne 0,\\
             \end{cases}
\eeqa
where again, in the second line above, we have omitted some inessential constants.

\subsection{General 2-point planar irregular graphs}

Let us now prove that in the case of a general 2-point planar irregular Feynman graph, the considerations of the previous subsection still hold.

\begin{figure}
\centerline{\epsfig{figure=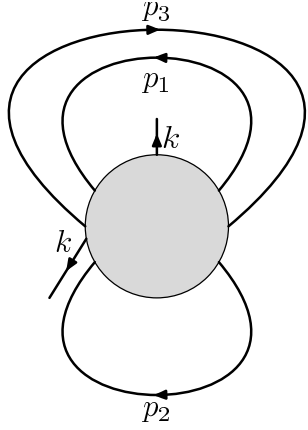,width=4cm} }
\caption{An example of a rosette obtained from a two-point graph with $3$ internal lines.}
\label{roz-2p}
\end{figure}

As before, let us denote the external moment of a two-point graph by
$k$. One has  $B=2$.  
 When shrinking the graph to a rosette (as described in section \ref{Feynman}), one
can always choose one of  the external legs to break the "external face". The other external leg then breaks 
an internal face (see for example Fig. \ref{roz-2p}, where this internal face 
corresponds to  line $p_1$). 

%Let us denote by $a^j p_j$ the respective liniar combination
% of the internal momentas for this internal face. 

The general amplitude contains an oscillating factor (see again \cite{filk} for further details)
$$ e^{-ik \wedge (a^j p_j)} .$$ The linear combination $a^j p_j$ corresponds to the sum of 
momenta of the lines arching over this second external leg, where the coefficients $a_j$ are $\pm 1$ . 
In the example of Fig. \ref{roz-2p}, because of the orientation of the lines 
one has to take the internal momenta $p_1$ and $p_3$ with opposite signs: $a^jp_j=p_1-p_3$.
A general Feynman amplitude writes
\beqa
{\cal A}_G= K \int \prod_{i=1}^L d^4 p_i \frac{e^{-i k \wedge (a^j
    p_j)}}{\prod_{i=1}^L (p_i^2 + m^2)}.
\eeqa
As in the case of the non-planar tadpole, we express now the integral above in
function of the Schwinger parameters $\alpha_\ell$ ($\ell =1,\ldots, L$). One
can then check (see \cite{param-GMRT} for a more detailed analysis) that after performing the Gaussian integrals over the
internal momenta $p_i$, one obtains an integral of the following type
\beqa
\label{int1}
\int_0^{\infty} 
U^{-2}(\alpha) e^{-m^2\sum_{j=1}^L\alpha_j^2 } 
e^{- \frac{P (\alpha)}{U(\alpha)} k^2}.\exp (- \frac{k \circ  k}{4 \sum_{{\rm overarching}\ \ell } \alpha_\ell})
\eeqa
%Let us now explain what we mean by these notations. 
where
\beqa
k\circ p = - k_\mu (\Theta^2)^{\mu \nu} p_\nu , \ (\Theta^2)^{\mu\nu}=\Theta^{\mu\rho}\Theta_{\rho}^\nu ,
\eeqa
$U(\alpha)$ and $P(\alpha) k^2 = V(\alpha, k)$ are the usual Symanzik (topological) polynomials
of the ordinary commutative $\phi^4$ theory (since we deal with a two-point
function there is a  single external invariant $k^2$ in factor in $V$).
 The sum $\sum_{{\rm overarching}\ \ell } \alpha_\ell$  above denotes the sum on the parameters $\alpha$ associated to the tree lines (which were reduced to obtain the rosette) as well as on the parameters $\alpha$ associated to the lines overarching the internal broken face ($\alpha_1+\alpha_3$ in the example of Fig. \ref{roz-2p}). 
Note that
\beqa
\label{grad}
({\rm deg}\, U)^2 = 2 (L-(n-1)).
\eeqa
Furthermore, since $L= 2n -N/2$ and we deal with the two-point
function ($N=2$) one has
\beqa
\label{grad1}
({\rm deg}\, U)^2 = 2 (L-(n-1))=L+1.
\eeqa
Let us recall that the polynomials $U$ and $V$ are explicitly positive.

Note that $\Theta^2=-\theta^2 {\rm Id}$
 so that integral \eqref{int1} writes
\beqa
\label{inte1}
\int_0^{\infty}  U^{-2}(\alpha) e^{-m^2\sum_{j=1}^L\alpha_j^2 } 
e^{- \frac{P (\alpha)}{U(\alpha)} k^2} \exp (- \frac{\theta^2 k^2}{4 \sum_{{\rm overarching}\ \ell } \alpha_\ell}) 
 \prod_{i=1}^L d \alpha_i. 
\eeqa
In the uv regime ($\alpha_\ell \to 0$), one needs to consider only 
\beqa
\label{int}
\int_0^{\infty}  U^{-2}(\alpha) e^{-m^2\sum_{j=1}^L\alpha_j^2 } 
 \exp (- \frac{\theta^2 k^2}{4 \sum_{{\rm overarching}\ \ell } \alpha_\ell}) 
 \prod_{i=1}^L d \alpha_i    .
\eeqa
We now analyze this integral by performing the following
change of variable:
\beqa
\label{change}
\alpha_i = k^2 \alpha'_i, \ i=1,\ldots , L. 
\eeqa
Using \eqref{grad1}, the integral \eqref{int} becomes 
\beqa
\label{int2}
\frac{1}{k^2}\int_0^{\infty} U^{-2}(\alpha') e^{-m^2\sum_{j=1}^L\alpha_j^{'2} } 
\exp (- \frac{\theta^2 }{4 \sum_{{\rm overarching}\ \ell } \alpha'_\ell})  \prod_{i=1}^L d \alpha'_i  .
\eeqa
This remaining integral is convergent. We have thus established the asymptotic behavior of 
${\cal A}_G$ to be $\frac{const}{k^2}$ at small $k$, as for the "non-planar" tadpole. Moreover when
$\theta \to 0$ we recover easily on (\ref{int1}) the usual mass and wave-function divergences associated to $G$.

\section{The noncommutative model and its renormalization} \label{sectionmodel}
\renewcommand{\theequation}{\thesection.\arabic{equation}}   
\setcounter{equation}{0}

The translation invariant $\phi^{\star 4}_4$  model defined in \cite{noi} has action
\beqa
\label{revolutie}
S_\theta [\phi]=\int d^4 p (\frac 12 p_{\mu} \phi  p^\mu \phi  +\frac
12 m^2  \phi  \phi   
+ \frac 12 a  \frac{1}{\theta^2 p^2} \phi  \phi  
+ \frac{\lambda }{4!} V_\theta ),
\eeqa
where  $a $ is  some dimensionless parameter.
The propagator is
\beqa
\label{propa-rev}
C(p) = \frac{1}{p^2+\mu^2+\frac{a}{\theta^2 p^2}} \, .
\eeqa
We further choose $ a \ge 0 \, $
so that this propagator is positively defined.

Let us now recall from \cite{noi} the following table summarizing the renormalization of the model \eqref{revolutie}:
\begin{center}
\begin{tabular}{|l|c|c|}\hline
%& \multicolumn{2}{|c|}{{\it the model \eqref{revolutie} }}\\ 
%\hline
 &  2-point function & 4-point function\\
\hline
planar regular & renormalization (mass and wave function) & renormalization ($\lambda$)\\
\hline
planar irregular & finite renormalization ($a$) & convergent \\
\hline
non-planar  & convergent & convergent \\
\hline
\end{tabular}
\end{center}
%where ren means renormalizable.and conv means convergent.

\section{The commutative limit} \label{sectionlimit}
\renewcommand{\theequation}{\thesection.\arabic{equation}}   
\setcounter{equation}{0}

In the limit $\theta\to 0$, equations \eqref{revolutie} or \eqref{propa-rev}  
no longer make sense.
This phenomenon takes place because the limit is done in a too direct, ``naive'' way. 
One should proceed as indicated by the analysis of the previous section.
As we have seen, when $\theta \to 0$ the convergent integrals of planar irregular two point graphs
become the usual divergent ones responsible for ``some part'' of the mass 
and wave function renormalizations. Furthermore, recall that the integrals of non-planar graphs also become divergent when $\theta\to 0$.   
We thus propose the following action with ultraviolet cutoff $\Lambda$
\beqa
\label{act-limita}
S_{\Lambda, \theta} [\phi]&= \int d^4 p \left[\frac 12  \eta^{-1}(p/\Lambda) p^2 \phi^2 + \frac 12 m^2 \phi^2 +
\frac{\lambda}{4!} V_\theta 
+ \frac 12 \delta_{Z'} p^2 \phi^2  + \frac 12 \delta_{Z"} (  1-T(\Lambda, \theta))
p^2 \phi^2  \right.  \nonumber\\ 
&
 + \frac 12 \delta_{m'}\phi^2
\left.+\frac 12 \delta_{m''} \phi^2  (1-T(\Lambda, \theta))
+\frac 12 \delta_a \frac{1}{\theta^2 p^2} \phi^2 T(\Lambda, \theta)
+\frac 12 a \frac{1}{\theta^2 p^2} \phi^2 T(\Lambda, \theta)\right.\nonumber\\
&\left. 
+\frac 12 \delta_{m'''} \phi^2  (1-T(\Lambda, \theta))
+\frac{\delta_{\lambda'}}{4!} (1-T(\Lambda, \theta))V_\theta
+\frac{\delta_{\lambda''}}{4!} T(\Lambda, \theta)V_\theta
\right],
\eeqa
where we have written the counterterms associated to \eqref{revolutie}. The cutoff $\Lambda$ is some ultraviolet scale with the dimension of a momentum. The function $\eta^{-1}(p/\Lambda) $ is a standard momentum-space ultraviolet cutoff which truncates momenta higher than $\Lambda$ in the propagator (\ref{propa-rev}). For that 
$\eta(p)$ could be a fixed smooth function with compact support interpolating smoothly
between value 1 for $\vert p/\mu \vert \le 1/2$ and 0 for $|p/\mu| \ge 1$.
Furthermore $T(\Lambda, \theta)$ is some smooth function satisfying the following conditions:
\beqa
\label{limite}
%T(x)={\mbox{lim}}_{t\to 0} (1- e^{-x/t}).
{\mbox{lim}}_{\theta \to 0} T(\Lambda, \theta)\frac{1}{\theta^2}&=& 0,\nonumber\\
{\mbox{lim}}_{\Lambda \to \infty} T(\Lambda, \theta)&=& 1.
\eeqa
There are of course infinitely many functions which satisfy these conditions, for instance a possibility is
\beqa
\label{candidat}
T(\Lambda, \theta)= 1- e^{-\Lambda^6 \theta^3},
\eeqa
(where the factor in the exponential has been chosen to be dimensionless).

Let us now comment on the action \eqref{act-limita}:
\begin{itemize}
\item the term $\delta_{Z'}$ corresponds to the planar regular graphs, while $\delta_{Z''}$ corresponds to the planar irregular and non-planar graphs;
\item the term $\delta_{m'}$ is the mass counterterm associated to the planar regular graphs, $\delta_{m''}$ is  associated to the planar irregular graphs and   $\delta_{m'''}$ is  associated to the non-planar graphs;
\item the term $\delta_a$ is the counterterm associated to the parameter $a$;
\item the term $\delta_{\lambda'}$ corresponds to the planar regular graphs, while $\delta_{\lambda''}$ corresponds to the planar irregular and non-planar graphs;
\end{itemize}
Note that the $\delta_a$ term come into the picture only when $\theta \ne 0$ (thanks to the $T$ function). Thus the $T$ function introduced here switches between the mass counterterm $\delta_{m''}$ (which is not present when $\theta\ne 0$ and the $\delta a$ counterterm. It it the main ingredient of the mechanism we propose here.

By taking the limit $\theta \to 0$, one obtains, using \eqref{limite} the usual commutative action
\beqa
\label{limita1}
{\mbox{lim}}_{\theta \to 0} S_{\Lambda, \theta}= S_{\Lambda}
=\int d^4 p 
[\frac 12  p^2 \phi^2 + \frac 12 m^2 \phi^2 +
\frac{\lambda}{4!} V
+ \frac 12 \delta_Z p^2 \phi^2 + \frac 12 \delta_{m}\phi^2 
+\frac{\delta_\lambda}{4!}V],
\eeqa
where we have rewritten
\beqa
\label{suma}
\delta_{m'}+\delta_{m''}+\delta_{m'''}=\delta_m
\eeqa
as the usual total mass counterterm and $V$ is the interaction potential obtained in the commutative limit ({\it i. e.} when the non-commutative Moyal-Weyl product simply becomes the usual pointlike multiplication of fields). Analogous relations as \eqref{suma} also hold for $\delta_Z$ and $\delta_\lambda$. 
%Note however that the difference between these three parameters is that, in the case of $\delta_m$, some part of it (namely $\delta_{m''}$) is switched to the counterem $\delta_a$ when $\theta\ne 0$.
 
Let us now make a few remarks. First, we recall that in \cite{param-GMRT} was proved that the convergence of the non-planar graphs improves proportionally to the value of their genus $g$. We don't take into consideration here this phenomenon, since what we are interested in is to show a mechanism which will transform these graphs which diverge when $\theta\to 0$, without looking into detail of the degree of convergence of the ``initial'' graph ({\it i.e.} when $\theta\ne 0$).

Indeed, when $\theta\ne 0$, the counterterms $\delta_{Z''}, \delta_{m''}, \delta_{m'''}$ and $\delta_{\lambda''}$ no longer survive (as requested by the renormalization results recalled in the previous section). When the parameter $\theta$ is switched off, all these counterterms come back to life and sum up in relations of type \eqref{suma}.

Let us end this paper by concluding that {\it the limit mechanism proposed here has nothing arbitrary, but is simply dictated in a natural way by the behavior of the $2-$point planar
irregular Feynman amplitudes} (as studied in section \ref{2puncte}).

%One can now use again \eqref{limite} to show the following result:
%\beqa
%\label{limita2}
%{\mbox{lim}}_{\Lambda \to \infty} S_{\Lambda, \theta}= S_{\theta}.
%\eeqa
%Therefore taking first the limit $\Lambda \to \infty$ at fixed $\theta$
%we get the renormalized theory \cite{noi}  Taking first the limit
%$\theta \to 0$ we get the ordinary commutative $\phi^4_4$ theory with cutoff $\Lambda$. Then taking
%$\Lambda \to \infty$ recovers the renormalized  $\phi^4_4$ without cuitoff (in the perturbative sense since this theory
%is not known to exist nonpertutrbatively). But we can now also study double limits 
%$\Lambda \to \infty$ and $\theta \to 0$ in a coupled way. If $\theta \to 0$ much faster than $\Lambda \to \infty$
%it is clear that we can get as a limit the renormalized ordinary $\phi^4_4$ theory. But 
%depending on the function $T$ and the coupled rate at which  $\theta \to 0$ and $\Lambda \to \infty$
%we may also find regimes which could interpolate between the theory \cite{noi} 
%and the the renormalized ordinary $\phi^4_4$ theory. Remark however that for phenomenolgical applications
%we may perfectly well work with fixed ultraviolet cutoff $\Lambda$ and let $\theta \to 0$, as we are interested
%in understanding the transition from noncommutative to commutiave field theory, not a more fundamental 
%theory without any ultraviolet cutoff.

\bigskip
{\bf Acknowledgment:} 
The authors warmly acknowledge Axel de Goursac for intensive discussions. A. Tanasa was partially supported by the CNCSIS Grant ``Idei''.

\end{document}